\documentclass{article}

\usepackage{hyperref}
\usepackage[numbers]{natbib}
\newcommand{\Schw}{\Big(1-\frac{2M}{r}\Big)}
\newcommand{\schw}{(1-2M/r)} 
\newcommand{\eroot}{\sqrt{e^2-1+\frac{2M}{r}}}
\newcommand{\erootinline}{\sqrt{e^2-1+2M/r}}

\usepackage{amsmath,amsfonts,graphicx} 

\begin{document}

\title{Schwarzschild spacetime under\\generalised Gullstrand-Painlev\'e slicing}
\author{Colin MacLaurin\footnote{colin.maclaurin@uqconnect.edu.au}\\University of Queensland, Australia}
\date{November 2019}
\maketitle

\begin{abstract}
We investigate a foliation of Schwarzschild spacetime determined by observers freely falling in the radial direction. This is described using a generalisation of Gullstrand-Painlev\'e coordinates which allows for any possible radial velocity. This foliation provides a contrast with the usual static foliation implied by Schwarzschild coordinates. The $3$-dimensional spaces are distinct for the static and falling observers, so the embedding diagrams, spatial measurement, simultaneity, and time at infinity are also distinct, though the $4$-dimensional spacetime is unchanged. Our motivation is conceptual understanding, to counter Newton-like viewpoints. In future work, this alternate foliation may shed light on open questions regarding quantum fields, analogue gravity, entropy, energy, and other quantities. This article is aimed at experienced relativists, whereas a forthcoming series is intended for a general audience of physicists, mathematicians, and philosophers.
\end{abstract}



\section{Introduction}
Schwarzschild-Droste spacetime is most commonly expressed in terms of Schwarzschild-Droste coordinates:
\begin{equation}
\label{eqn:Schwcoords}
ds^2 = -\Schw dt^2+\Schw^{-1}dr^2+r^2d\Omega^2
\end{equation}
where $d\Omega^2 := d\theta^2+\sin^2\theta\,d\phi^2$. This is a natural choice because of their simplicity and intuition, also $r$ is the curvature coordinate and $\partial_t$ is the preferred Killing vector field. However in pedagogical settings, many presentations express the properties of space and time in terms of the foliation $t=\textrm{const}$, but without a clear qualifier about this choice. This downplays the lesson that space and time are relative to the observer. This is particularly true in the case of the radial proper distance, and also for the ``time at infinity'' when interpreted as a global simultaneity convention. We seek to improve conceptual understanding by taking a complementary description based on freely-falling observers, and re-examining familiar descriptions from this alternate $3+1$-splitting. These observers and their space (the hypersurfaces orthogonal to them) are conveniently described using a generalisation of Gullstrand-Painlev\'e coordinates which allows for all possible radial velocities.

While most textbook material and research calculations correctly account for coordinates or frames, some conceptual confusion remains. For example the mathematics of arbitrary foliations is well understood and clearly taught, including the intrinsic and extrinsic curvature of hypersurfaces, and the ADM formalism along with lapse-shift notation. However this technical knowledge has not always been utilized in some expositions of Schwarzschild coordinates. Another topic which is generally handled well is the careful extraction of measurable observables, for instance the decomposition of the velocity gradient of a congruence into shear, expansion, and vorticity. Yet while for proper distance the general definition $\int ds$ is standard, it is rarely applied to give anything beyond the static radial measurement $(1-2M/r)^{-1/2}dr$. The book by \citet{taylorwheeler2000} is a notable exception. Yet it seems ``neo-Newtonian'' \citep{eisenstaedt1989low} interpretations of relativity have not completely died out.

The generalised Gullstrand-Painlev\'e coordinates are well suited to the falling observers, in that the $x^0$-coordinate is their proper time. These coordinates are convenient for computation and intuition. One might protest that calculations can be made in any coordinate system, however in history new coordinates have often helped to advance understanding \citep{liberati+2018}. In particular, the popularisation of Eddington-Finkelstein and Kruskal-Szekeres coordinates extended understanding across the horizon(s). However these null coordinates are less insightful for timelike observers \citep{finch2015}. There are probably hundreds of coordinate systems for Schwarzschild spacetime used in the literature \citep{vilain1992} \citep[\S7]{synge1960} \cite[\S2.2]{muellergrave2009} \citep{smarryork1978} \citep[\S10]{gourgoulhon2012}, but the coordinates studied here are arguably the most intuitive for radial timelike geodesics. Another obvious ``natural'' choice is circular orbits, however these only exist for $r>3M$ so cannot probe the horizon.


In future work, the generalised Gullstrand-Painlev\'e coordinates could have applications to open research questions about any quantities which depend on foliation, such as entropy or the decomposition of quantum fields on curved spacetime into positive and negative frequency modes \citep{wald1994}. A special case has given alternate descriptions for Hawking radiation \citep[\S3]{krauswilczek1994} \citep{parikhwilczek2000} \citep[\S4.3]{belgiorno+2018}. Also related coordinates are considered in laboratory analogues of gravity based on fluid flow \citep{rosquist2009} \citep{liberati+2018}.

Section~\ref{sec:embedding} presents isometric embedding diagrams for the new spatial slices which are orthogonal to the observer congruence. This is a way of visualising the curvature of $3$-dimensional space. Section~\ref{sec:spatialdistance} discusses measurement of the radial proper distance, for different observers including inside the horizon. Section~\ref{sec:spacetimecoords} clarifies some potential misconceptions about coordinate basis vectors and coordinate gradients. While this section contains fairly straightforward material, it has rarely been explained in print. Section~\ref{sec:simultaneity} explains the concept of ``time at infinity'' contains an implicit simultaneity convention, if interpreted globally. The falling observers offer a different simultaneity convention, which helps to avoid recurring misconceptions. The following Sections~\ref{sec:radialgeodesics} and \ref{sec:gengpcoords} give a careful analysis of the worldlines and coordinate charts; for a quicker read, glance at Figure~\ref{fig:rainhaildripmist} and Equation~\ref{eqn:GenGPelement} before skipping to Section~\ref{sec:embedding}.


\section{Radial timelike geodesics}
\label{sec:radialgeodesics}

The geodesics of Schwarzschild (and Kerr) have been thoroughly researched. Still, it is helpful to clarify the allowable radial timelike geodesics, in particular the existence of negative Killing energy orbits within the extended manifold. Textbooks justifiably focus on orbits with angular momentum. Of our four classes of radial geodesics given below, \citet[\S19]{chandrasekhar1983} covers just one in detail (but clearly implies there are others), \citet[\S2.3.2]{frolovnovikov1998} briefly allude to all four, and \citet[\S4.10]{oneill1995} has a detailed study for Kerr; however the older classic \citet[ff.]{hagihara1970} may exclude the $r<2M$ black hole region and hence the negative Killing energy geodesics, and the language in \citet{taylorwheeler2000} denies their existence.

\begin{figure}
\label{fig:rainhaildripmist}
\centering
\includegraphics[width=0.75\textwidth]{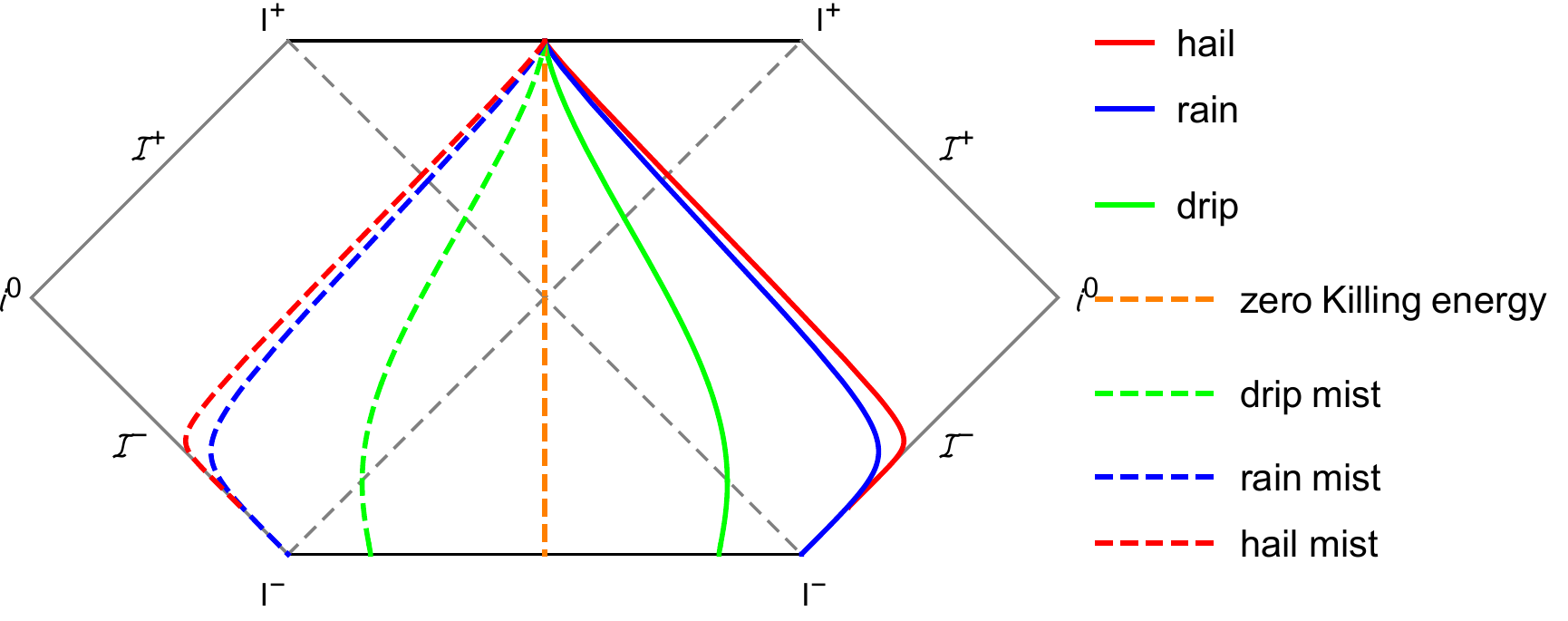}
\caption{Some maximally extended geodesics in a Penrose diagram, picked to coincide at the ``event'' $(t,r)\rightarrow(0,0)$. Rain and hail start from past timelike infinity, and fall inwards to the black hole singularity; their mist variants are similar. Drip and drip mist start from the past singularity in the white hole interior, fall out of the past horizon into an exterior region, then reach a maximum height before falling through the future horizon into the black hole interior. Zero Killing energy geodesics pass from the white hole interior through the bifurcate horizon directly into the black hole interior. From our universe, negative energy worldlines may be achieved by passing inside the horizon and then accelerating ``outwards''.}
\end{figure}

The quantity $e$ derived from the ``static'' Killing vector field $\boldsymbol\xi$ which is timelike at infinity, is a natural parameter choice:
\begin{equation}
e := -\mathbf u \cdot \boldsymbol\xi
\end{equation}
where $\mathbf u$ is the $4$-velocity field, and the dot represents the metric. This Killing energy per mass or ``energy per mass at infinity'' is preserved along geodesics.\footnote{Infinity can be made rigorous by conformal compactification, which produces a new manifold with a boundary consisting of timelike, null, and spacelike infinities. However for our purposes a simple limit $r\rightarrow\infty$ is often sufficient, or at least an approximation $r\gg 2M$. Physically, infinity is loosely analogous with Solar System observers far from a black hole.} $\boldsymbol\xi$ has the components $(1,0,0,0)$, where all components are expressed in Schwarzschild charts unless otherwise stated. (This is familiar and simple, but internal computations are made in Gullstrand-Painlev\'e or Kruskal-Szekeres coordinates where necessary.) The $4$-velocity is
\begin{equation}
\label{eqn:velocity}
u^\mu = \bigg(e\Big(1-\frac{2M}{r}\Big)^{-1},\pm\sqrt{e^2-1+\frac{2M}{r}},0,0\bigg)
\end{equation}

These have zero angular momentum, as determined from the Killing vector fields orthogonal to $\boldsymbol\xi$. 
\citet[\S B2]{taylorwheeler2000} use the metaphors \emph{hail}, \emph{rain}, and \emph{drips} for classes of ingoing radial motion, see Table~\ref{tbl:rainhaildripmist}. Rain are the quintessential radial geodesics, and ``fell from rest at infinity'' so to speak, or have $e=1$ to be precise. Hail has $e>1$, and fell from infinity with initial inward velocity having Lorentz factor $\gamma=e$ relative to a static observer there. Drips have $0<e<1$ and fell from rest at some finite $r_\textrm{max}>2M$. Taylor \& Wheeler state these ``cover all possible radially moving free-float frames'' \citep[\S B2]{taylorwheeler2000}, however there are trajectories with $e\le 0$ which exist only inside the horizon. We dub these \emph{mist} because they fall more slowly than drips (in the sense their relative $3$-velocity points outwards), and seem ethereal to observers in region I. See Table~\ref{tbl:rainhaildripmist}. To achieve these worldlines, pass into the horizon, accelerate sufficiently ``outwards'', then return to freefall. This possibility of negative Killing energy observers is best known in the context of the Penrose process or superradiant scattering in Kerr spacetime, or from the heuristic description of Hawking radiation as particle pairs \citep[\S1]{hawking1975}. It can occur when the relevant timelike Killing vector field becomes spacelike.

\begin{table}
\label{tbl:rainhaildripmist}
\centering
\begin{tabular}{|c|c|c|c|}
\hline
Nickname & Traditional name & Killing energy per mass & Range \\
\hline
hail & hyperbolic & $e>1$ & all \\
rain & parabolic & $e=1$ & all \\
drip & elliptic & $0<e<1$ & $r \le r_\textrm{max} = \frac{2M}{1-e^2}$ \\
mist & \rule{0.7cm}{0.5pt} & $e\le 0$ & $r<2M$ \\
\hline
\end{tabular}
\caption{Radial timelike geodesics in terms of $e$. This is restricted to regions I and II. Note the energy measured by any \emph{local} observer is always positive.}
\end{table}

So far this discussion is limited to inward motion in the ``physical'' spacetime consisting of black hole interior and one exterior region. In the maximal analytic extension of the manifold, the $e<0$ worldlines emerge from the parallel exterior region, if continued backwards as geodesics, as Figure~\ref{fig:rainhaildripmist} shows. We can also subdivide ``mist'' into rain mist ($e=-1$), hail mist ($e<-1$), drip mist ($-1<e<0$), and zero Killing energy observers ($e=0$). The sign of $dr/d\tau$ in Equation~\ref{eqn:velocity} is an additional parameter which specifies ingoing or outgoing motion (lower and upper signs respectively), which extends the classes to outgoing variants, where allowed. Table~\ref{tab:regions} shows the allowed parameter combinations for all $4$ regions. Hence the extended) timelike radial geodesics are classified uniquely, modulo translation in ``time'' ($\boldsymbol\xi$).

\begin{table}
\label{tab:regions}
\centering
\begin{tabular}{|c|c|c|}
\hline
Energy & Direction & Regions \\
\hline
$e>0$ & ingoing & I and II \\
$e<0$ & ingoing & III and II \\
$e>0$ & outgoing & IV and I \\
$e<0$ & outgoing & IV and III \\
$e=0$ & \rule{0.7cm}{0.5pt} & IV and II \\
\hline
\end{tabular}
\caption{Allowed regions, this generalises Table~\ref{tbl:rainhaildripmist}. Drip and drip mist are further subject to $r\le r_\textrm{max} = 2M/(1-e^2)$. There is no ingoing/outgoing freedom for $e=0$.}
\end{table}


\section{Generalised Gullstrand-Painlev\'e coordinates}
\label{sec:gengpcoords}

For $e\ne 0$ we use a coordinate $T \equiv T_e$ which is proper time along the worldlines. While one possibility is to extend a single local frame outwards by geodesics, we consider an entire congruence of worldlines \citep[\S2.6]{malament2007} \citep[\S3.2.2]{frolovnovikov1998}. The metric becomes
\begin{equation}
\label{eqn:GenGPelement}
ds^2 = -\frac{1}{e^2}\Schw dT^2 \mp\frac{2}{e^2}\eroot dT\,dr+\frac{1}{e^2}dr^2+r^2d\Omega^2
\end{equation}
This line element is suited to the congruence, and aids intuition and computation. (We avoid the term ``adapted'' coordinates which implies $g_{0i}=0$.) The transformation from Schwarzschild coordinates is
\begin{equation}
\label{eqn:dT}
dT = e\,dt \mp\Schw^{-1}\eroot\,dr
\end{equation}
Note this must match $\partial T/\partial t\cdot dt + \partial_T/\partial r\cdot dr$. The rain case $e=1$ is due to \citet{gullstrand1922} and \citet{painleve1921}. It was generalised by \citet{gautreauhoffmann1978} using the parameter $r_\textrm{max}$ described previously, hence limited to the drip case $0<e<1$. \citet{martelpoisson2001} derived a similar coordinate $eT$, using instead a parameter $p:=1/e^2$ they limited to the rain and hail cases, and under which Eddington-Finkelstein null coordinates are a limiting case. \citet{finch2015} clarified the unity for $e>0$. \citet{bini+2012} have the most general treatment, using $e$ as parameter, including outgoing motions, and applied to various spacetimes. There are numerous other related works. My contribution is to clarify the allowable worldlines in this context (specifically $e\le 0$), and to explore the resulting properties of the $3+1$-splitting.

Various related derivations have been given, which can be generalised if necessary. One uses local Lorentz boosts from the static to falling orthonormal frames \citep[\S B-4]{taylorwheeler2000} \citep{linsoo2013}. Another considers the proper time on freely falling clocks \citep{gautreauhoffmann1978} \citep[\S15]{moore2012}. We follow instead a mathematically elegant approach which defines the time gradient from the co-velocity:
\begin{equation}
\label{eqn:dTdualvelocity}
dT := -\mathbf u^\flat
\end{equation}
where $\mathbf u^\flat$ is dual to the $4$-velocity $\mathbf u$, and the minus sign compensates for our signature convention -+++. This definition is the unique choice which is both proper time along the worldlines: $dT/d\tau\equiv dT(\mathbf u)=1$, and constant along the $3$-space orthogonal to them --- that is, Einstein simultaneous: $dT(\mathbf v)=0$ for $\mathbf{u}\cdot\mathbf{v}=0$.\footnote{Recall the intuition for combining a vector with a $1$-form $dT$ to produce a scalar, as the number of level sets $T=\textrm{const}$ crossed by the vector \citep[\S3.3]{schutz2009} \citep[\S2.5]{misner+1973}} (Equation~\ref{eqn:dTdualvelocity} is a valid definition locally if and only if a timelike congruence is \emph{geodesic} and \emph{vorticity-free}. With the inclusion of a scalar integrating factor $1/N$, where $N$ is the lapse, the geodesic requirement is dropped.) This co-velocity approach was applied in our context by \citet{martelpoisson2001}, \citet{finch2015}, and \citet{bini+2012}. Amongst general treatments, \citep[\S4.6.2]{ellis+2012} and \citep[\S2.3.3]{poisson2004} are especially clear and relevant here; see also early sources \citep{synge1937} \citep[\S2.2]{ehlers1961} or numerical relativity textbooks. The vorticity-free requirement follows from Frobenius' theorem, see also \citep[\S2.3]{sachswu1977} \citep[\S B-3]{wald1984} \citep[\S2.12]{defeliceclarke1990} or the differential geometry literature.

The coordinates are regular at $r=2M$. The cross-term indicates motion relative to $r$, as seen from the inverse metric component
\begin{equation}
g^{Tr} = dT\cdot dr = dr(dT^\sharp) = dr(-\mathbf u) = -u^r = -\frac{dr}{d\tau} \ne 0
\end{equation}
in general, using Equations~\ref{eqn:velocity} and \ref{eqn:dTdualvelocity}. The coordinates have lapse $N=1$ since $T$ is proper time, also a shift of $(\mp\erootinline,0,0)$. Indeed, because of the unit lapse, the inverse metric components $(g^{Ti})_{i=1,2,3}$ are precisely the shift vector \citep[\S2]{bini+2012} \citep[\S21.4]{misner+1973}. The line element is independent of $T$, so unchanged by translation in $\partial_T$ which is the Killing vector field $\boldsymbol\xi/e$. Even though the normal $\mathbf u$ to the hypersurfaces has a nonzero $r$-component, the shift $4$-vector compensates for this, so the sum of these (normal times lapse plus shift) is the coordinate vector $\partial_T$. The $T=\textrm{const}$ slices are identical, it does not matter the symmetry underlying this is spacelike inside the horizon.

\begin{figure}[h]
\label{fig:GPregion}
\centering
\includegraphics[width=0.6\textwidth]{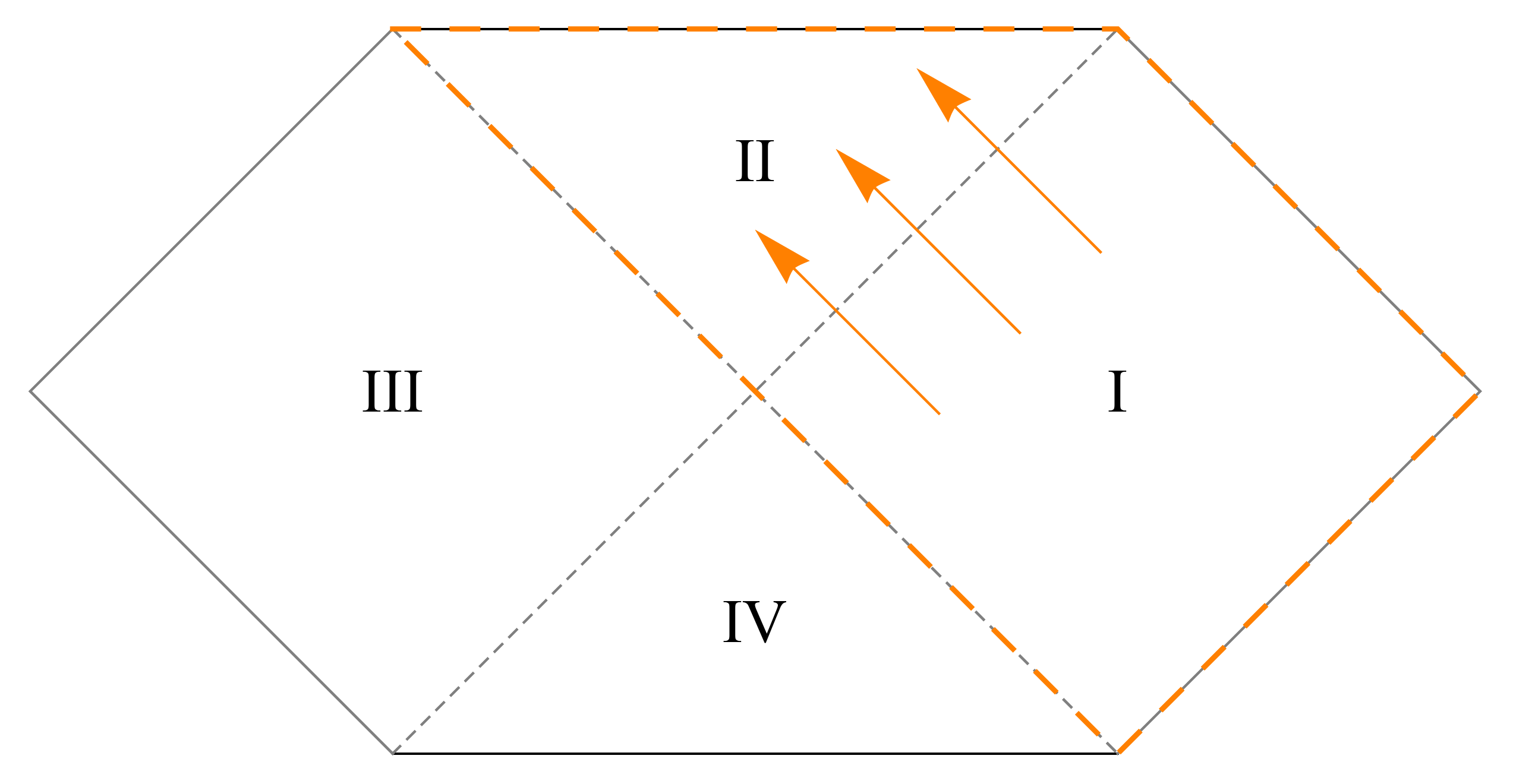}
\caption{Standard Penrose diagram for Schwarzschild spacetime, with regions dubbed our universe exterior (I), black hole interior (II), parallel universe exterior (III), and white hole interior (IV). The orange line borders our default generalised Gullstrand-Painlev\'e chart, for an ingoing variant with $e\ge 1$.}
\end{figure}

A single coordinate chart covers two adjacent regions of the extended manifold, including the horizon segment between them. Figure~\ref{fig:GPregion} shows an ingoing chart with $e>0$. The outgoing variant with $e>0$ covers regions IV and I. As pointed out for the $e=+1$ coordinates by \citep[\S2]{krauswilczek1994}, all regions can be covered by reversing time for these charts: this means nothing more exotic than reinterpreting $T$ as \emph{decreasing} towards the future. Hence in the reversed charts, \emph{minus} $\mathbf u$ is future-pointing. Our inclusion of $e<0$ coordinates allows a more intuitive interpretation: time-reversing a chart is equivalent to reversing both the sign of $e$ and ingoing / outgoing, as seen from Equation~\ref{eqn:velocity}. Table~\ref{tab:regions} summarises the variants. (Note the line element is formally the same under $\pm e$, but the meaning of $T$ is not.)

It must be qualified the drip and drip mist charts are bounded by $r < r_\textrm{max}$. \citet{martelpoisson2001} discount them for this reason, however the worldlines which motivate these charts only exist in the same sub-region, and we can certainly supplement with different coordinate charts. The boundary at $r_\textrm{max}$ is excluded because technically charts are open subsets; in fact the Christoffel symbol $\Gamma^T_{rr}$ diverges at the boundary. Any two complementary Gullstrand-Painlev\'e charts still omit one branch of the horizon. All four chart combinations together still omit the bifurcate horizon, as pointed out for Eddington-Finkelstein coordinates \citep{carter1973}. One can use Kruskal-Szekeres coordinates there. To a physicist these details seem pedantic, but they are important for the $e=0$ trajectories.

We briefly mention some properties useful in various $3+1$-applications. Firstly for the worldlines themselves, the kinematic decomposition shows shear and expansion, but zero vorticity and acceleration. The $3$-spaces $T=\textrm{const}$ have metric
\begin{equation}
^{(3)}ds^2 = \frac{1}{e^2}dr^2+r^2d\Omega^2
\end{equation}
in terms of $r$. (In fact this $3$-metric occurs within any static, spherically symmetric, vacuum spacetime \citep[\S3]{bini+2012}.) The Riemann tensor has (at most) one non-trivial component up to symmetry: $^{(3)}R^{\theta\phi}_{\hphantom{\theta\phi}\theta\phi}=(1-e^2)/r^2$, the Ricci tensor components $^{(3)}R^\theta_{\hphantom\theta \theta}$ and $^{(3)}R^\phi_{\hphantom\phi \phi}$ have the same value, and the Ricci scalar is double this \citep[\S3]{martelpoisson2001} \citep[\S3]{bini+2012}. Hence the curvature of $3$-space has the same sign everywhere, but not the same magnitude. The Kretschmann scalar is $4(1-e^2)^2/r^4$. The extrinsic curvature has trace $K$ behaving like $r^{-3/2}$ for $r\rightarrow 0$. This is significant for quantum fields on curved spacetime, where authors require ``nice slices'' with small intrinsic and extrinsic curvature \citep{lowe+1995} \citep{mathur2009exactly}. Another consideration is asymptotic flatness, but even the $e=\pm 1$ coordinates fail to meet some technical conditions here, so their ADM mass is not valid \citep[\S8.3.1]{gourgoulhon2012}. Hence our coordinates, though nice, are not nice enough at $r\rightarrow 0$ and $r\rightarrow\infty$ by some requirements. But they are physically motivated, unlike the contrived singularity-avoiding slices in the quantum field theory references, so might still prove useful even in these ongoing fields of research.


An additional coordinate system suited to $e\ne 0$ observers will also prove useful. Keep the new $T$-coordinate, but replace $r$ with $\rho\equiv\rho_e$ defined such that the metric becomes diagonal:
\begin{equation}
\label{eqn:lemaitrecoords}
ds^2 = -dT^2 + \frac{1}{e^2}\Big(e^2-1+\frac{2M}{r}\Big)d\rho^2 + r^2d\Omega^2
\end{equation}
A falling observer with the same $e$ as the coordinate system is comoving: $\rho=\textrm{const}$. Hence these coordinates are even better adapted to the congruence, at the expense of a less intuitive radial coordinate. These coordinates were given by \citet[\S4]{gautreauhoffmann1978} for $0<e<1$. They are a generalisation of the $e=1$ ``Lema\^itre coordinates'', but a special case of the general Lema\^itre[-Tolman-Bondi] coordinates for a spherically symmetric dust spacetime \citep{lemaitre1932}. The transformation from Schwarzschild coordinates is
\begin{equation}
d\rho = e\,dt \mp e^2\Schw^{-1}\Big(e^2-1+\frac{2M}{r}\Big)^{-1/2}dr
\end{equation}
While the line element Equation~\ref{eqn:lemaitrecoords} is formally the same for $\pm e$ and ingoing/outgoing variants, the coordinates $T$ and $\rho$ depend intrinsically on these.

Finally for $e=0$ observers (at $r\ne 2M$), ordinary Schwarzschild coordinates are adapted, because $r$ is \emph{purely} timelike to them ($dr\propto\mathbf u$) and $t$ purely spacelike, so each worldline has $t=\textrm{const}$. At the bifurcate horizon, Kruskal-Szekeres coordinates will suffice.

\section{Embedding diagrams: cones and funnels}
\label{sec:embedding}

\begin{figure}
\label{fig:embeds}
\centering
\begin{minipage}[b]{0.4\textwidth}
\includegraphics[width=\textwidth]{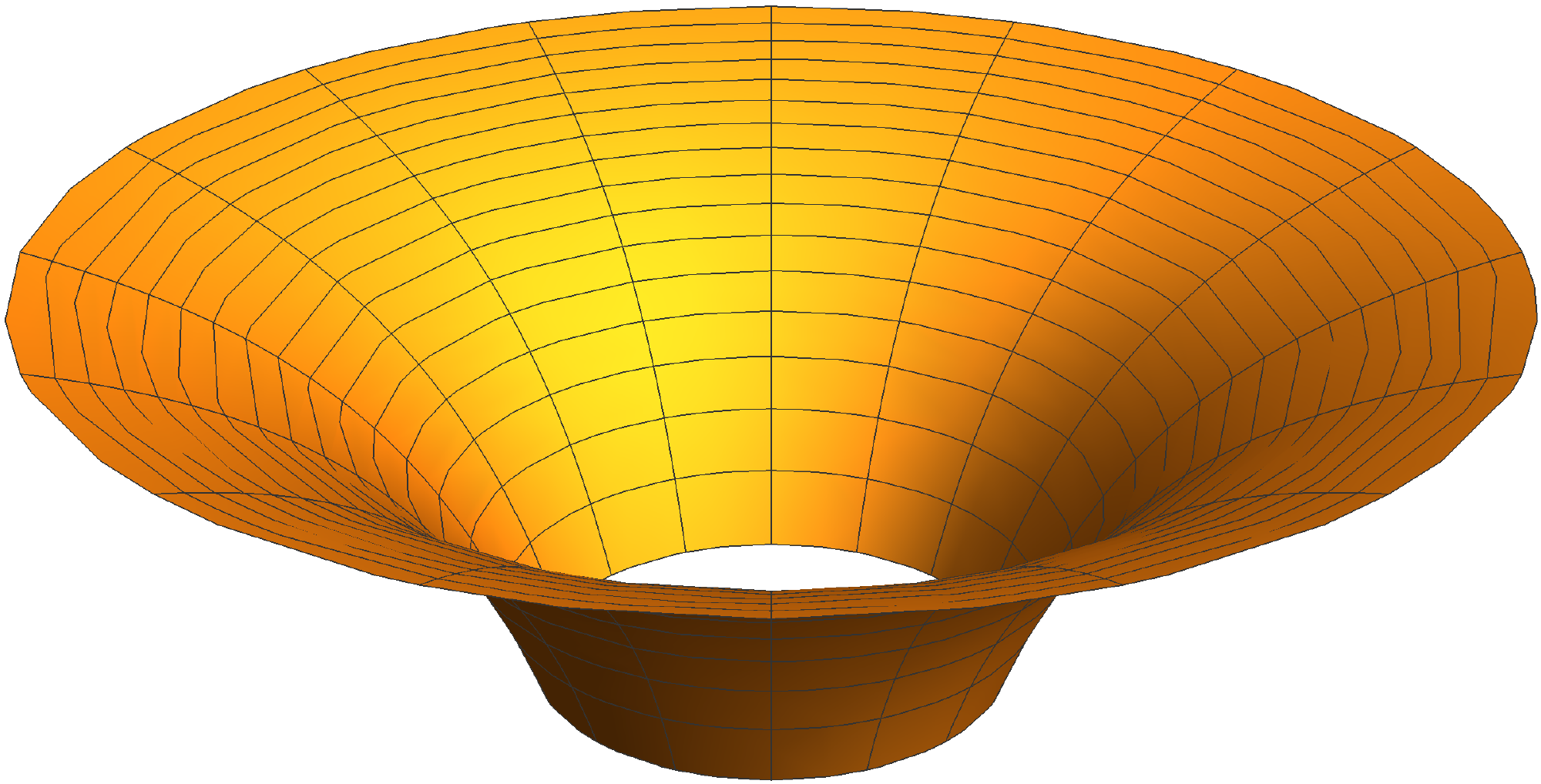}
\end{minipage}
\begin{minipage}[b]{0.4\textwidth}
\includegraphics[width=\textwidth]{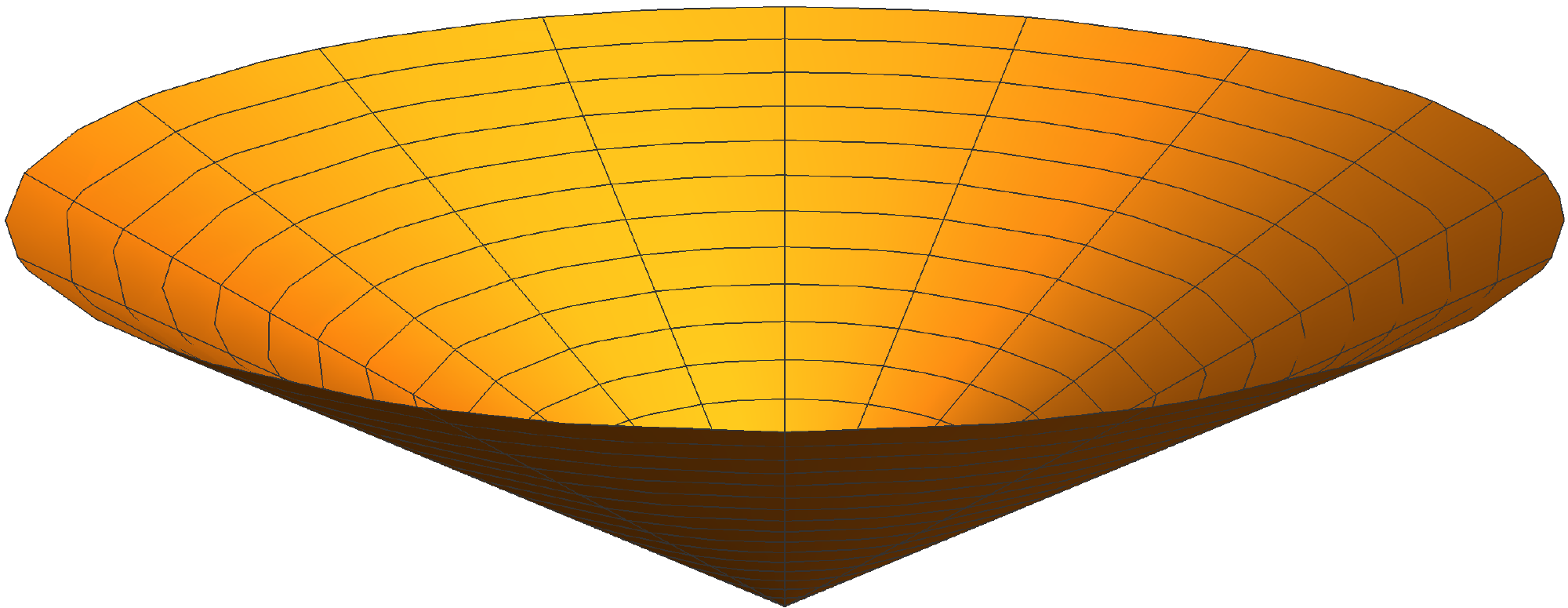}
\end{minipage}
\caption{Embedding diagrams in flat space, restricted to regions I and II only. The left is Flamm's paraboloid, which shows the curvature of static slices. The right diagram shows the curvature of the falling observer space $T=\textrm{const}$. The latter is a cone with $z\propto r$, and extends inside the horizon (with the tip excluded since $r=0$ is not part of the $4$-manifold). While a $2$-cone is intrinsically flat, the ``$3$-cone'' it represents is not.}
\end{figure}

One use of isometric embeddings is as a visual representation of curvature. For any foliation respecting the ``time'' and spherical symmetries, a $2$-dimensional equatorial surface $\theta=\pi/2$ of a constant time slice is representative. Under the usual static foliation $t=\textrm{const}$, the surface has metric
\begin{equation}
^{(2)}ds^2 = \Schw^{-1}dr^2+r^2d\phi^2
\end{equation}
This can be embedded in Euclidean $\mathbb R^3$: under cylindrical coordinates $(r,\phi,z)$, the surface $z=z(r)$ has metric
\begin{equation}
ds^2 = dr^2 + dz^2 + r^2d\phi^2 = \bigg(1+\big(\frac{dz}{dr}\big)^2\bigg)dr^2 + r^2d\phi^2
\end{equation}
The metrics match when $z=\sqrt{8M(r-2M)}$, a popular funnel-shaped surface of revolution known as ``Flamm's paraboloid'' \citep{flamm1916}. However this is not the only possible depiction of the curvature of space. The $T = \textrm{const}$ hypersurfaces have equatorial slice
\begin{equation}
^{(2)}ds^2 = \frac{1}{e^2}dr^2+r^2d\phi^2
\end{equation}
For drips and drip mist $0<|e|<1$, the embedded surface $z = \sqrt{1/e^2-1}\,r$ yields matching metrics, defined up to $r = r_\textrm{max}$. For $|e|=1$ the plane $z=0$ suffices, indeed many authors have noticed $3$-space is flat for $e=+1$ \citep{lemaitre1932} \citep{painleve1921}. For $|e|>1$ the above embedding approach fails, as the Euclidean surface must have $g_{rr}\ge 1$. However an alternate choice is to embed in Minkowski spacetime:
\begin{equation}
\label{eqn:embedMinkowski}
ds^2 = dr^2 - dz^2 + r^2d\phi^2 = \Big(1-\big(\frac{dz}{dr}\big)^2\Big)dr^2 + r^2d\phi^2
\end{equation}
This leads to $z = \sqrt{1-1/e^2}\,r$, as drawn in Figure~\ref{fig:embeds}.

The $e=0$ case is unique. These worldlines have $t=\textrm{const}$, and the hyperplanes orthogonal to them in the tangent space have $r=\textrm{const}$. In fact under this congruence, all events at a given $r=r_0$ in the same region (II or IV) mesh together to form a hypersurface $r=\textrm{const}$ on the manifold. Its metric follows from Equation~\ref{eqn:Schwcoords}:
\begin{equation}
^{(3)}ds^2 = \Big(\frac{2M}{r_0}-1\Big)dt^2 + r_0^{\hphantom{0}2}(d\theta^2+\sin^2\theta\,d\phi^2)
\end{equation}
which is the geometry $\mathbb R\times S^2$. The $2$-surface with $\theta=\pi/2$ embeds in Euclidean space as the cylindrical surface with $r = r_0$ and the $z$-axis replaced by $\sqrt{2M/r_0-1}\,t$. There is a single exception: at the bifurcate horizon the orthogonal hypersurfaces immediately exit the congruence, but if continued as spatial geodesics they extend into regions I and III as the static slices $t=\textrm{const}$. This is an Einstein-Rosen bridge, so the embedding diagram is two of Flamm's surfaces stuck together. Figure~\ref{fig:zerosimul} shows these hypersurfaces in a Penrose diagram.

Flamm's paraboloid is not the only depiction of the curvature of space. While the static foliation is a particularly natural choice, the $e=1$ choice is also natural and we argue more physically reasonable. Pedagogically, sources could more clearly qualify that Flamm's paraboloid is not the only possible embedding diagram, but represents the curvature of space as measured by observers at a fixed location. Having said that, many sources do provide embeddings from alternate slices, via $x^0 = \textrm{const}$ under various coordinates, or other creative choices \citep[\S21.8, \S31.6]{misner+1973} \citep{estabrook+1973} \citep{marolf1999} \citep[\S7.13]{hamilton2015}. Perhaps the Gauss-Codazzi equations for a congruence would sharpen the arguments about spatial geometry.

To push the point, we can make a ``fake black hole'' via an unusual foliation of Minkowski spacetime \citep{maclaurin2018fakeBH}. From spherical coordinates $-dt^2 + dr^2 + r^2d\Omega^2$, take an observer field moving radially as a function of $r$ only: $u^\mu = (u^t,u^r,0,0)$ where $u^t=\sqrt{1+(u^r)^2}$. A new time coordinate $dT := -\mathbf u^\flat/u^t$ has level sets $T = \textrm{const}$ orthogonal to the observers, even though an (Einstein-synchronised) proper time coordinate does not exist in general. The metric becomes
\begin{equation}
ds^2 = -dT^2 -2\frac{u^r}{u^t}dT\,dr +\frac{1}{1+(u^r)^2}dr^2 +r^2d\Omega^2  
\end{equation}
Various different choices of $u^r$ lead to $3$-slices which imitate selected aspects of static and falling observers in Schwarzschild spacetime, but here we duplicate Flamm's paraboloid by taking $u^r := \pm\sqrt{2M/(r-4M)}$. Since $g_{rr} \le 1$, (re-)embed in Minkowski spacetime as before in Equation~\ref{eqn:embedMinkowski}. The surface is valid for $r\ge 4M$, and formally matches Flamm's paraboloid! Admittedly this is contrived, as the worldlines are accelerated and seem unnatural, and the underlying $3$-geometries are distinct. Nevertheless, it cautions against overinterpretation of the funnel picture. Perhaps this foliation of Minkowski spacetime would make a useful test case for quantum effects in analogy with black holes, just as Rindler coordinates are; presumably here a null-result is expected.


\section{Spatial measurements}
\label{sec:spatialdistance}

In this section we discuss the spatial distance measured by observers of various velocities near a black hole. The familiar textbook ``radial proper distance'' interval
\begin{equation}
\label{eqn:staticdistance}
ds = \Schw^{-1/2}dr
\end{equation}
follows from setting $dt=d\theta=d\phi=0$ in Equation~\ref{eqn:Schwcoords}. But since the general definition of proper distance is $\int ds$ over any spacelike curve, we can apply the analogous procedure in the new coordinates, to measure along the slice $dT=0$:
\begin{equation}
\label{eqn:fallerdistance}
ds = \frac{1}{|e|}dr
\end{equation}
\citet{painleve1921} contrasted the above results in the $e=1$ case, and concluded general relativity is self-contradictory. Instead, these describe measurements by different observers: \emph{static} and \emph{falling} respectively.\footnote{We write $ds$ for both, but these should not be equated, as they are restrictions of the full spacetime metric along different $4$-vectors.} Equation~\ref{eqn:fallerdistance} is valid even inside the horizon! While the common interpretation of Equation~\ref{eqn:staticdistance} as measurement by an observer at infinity is not completely without merit, measurement by local observers is more directly meaningful. Any generalisations of this quantity are surprisingly little known. \citet{gautreauhoffmann1978} showed Equation~\ref{eqn:fallerdistance} for the drip case $0<e<1$, and \citet[\S B-3]{taylorwheeler2000} justify the rain case $e=1$. Similarly, the $3$-volume inside the event horizon for our congruence is $1/|e|$ times the Euclidean ball volume $\frac{4}{3}\pi(2M)^3$, which \citet{finch2015} showed for $e>0$.

There are other derivations of these results. The most fashionable conception of measurement uses clocks not rulers, because of the impossibility of Born-rigid objects. The Landau-Lifshitz radar metric \citep[\S79]{landaulifshitz1941} achieves this using null rays to probe nearby space, defining spatial distance using the time of their return journey. This leads to a $3$-metric $^{(3)}ds^2 = \gamma_{ij}dx^idx^j$ where
\begin{equation}
\label{eqn:radarmetric}
\gamma_{ij} := g_{ij} - \frac{g_{0i}g_{0j}}{g_{00}}
\end{equation}
in a given coordinate system. Since all null rays move at $c$ this might seem a preferred/absolute measure of spacetime, however it also depends on the motion and proper time of the radar device. Equation~\ref{eqn:radarmetric} presumes the device is comoving with the coordinate system: $x^i=\textrm{const}$, for $i=1,2,3$. In Schwarzschild coordinates the radar metric gives the $dt=0$ slice, and since it is the static observers which are comoving, the slice $dt=0$ including Equation~\ref{eqn:staticdistance} is the measurement of static observers. The generalised Gullstrand-Painlev\'e coordinates also have static observers as comoving, so yield the same radar metric. For the falling frames we require instead the Lema\^itre coordinates, for which the radar metric is the line element Equation~\ref{eqn:lemaitrecoords} with $dT=0$. This must equal the Gullstrand-Painlev\'e line element with $dT=0$, as both coordinates express the same $4$-metric tensor $\mathbf g$. In particular, falling observers measure the radial distance Equation~\ref{eqn:fallerdistance} as claimed.

Another approach uses the spatial projector tensor
\begin{equation}
\label{eqn:projector}
P_{\mu\nu} := g_{\mu\nu}+u_\mu u_\nu
\end{equation}
for a given observer $\mathbf u$, to derive its spatial metric $P_{\mu\nu}dx^\mu dx^\nu$ \citep[\S6.1]{defelicebini2010}. The radial distance is given by contracting over the radial vector, which naively would be the coordinate basis vector $\partial_r$, which picks out the $P_{rr}$-component. For static observers and Schwarzschild coordinates, $P_{rr}$ leads to Equation~\ref{eqn:staticdistance}, whereas for falling observers in Gullstrand-Painlev\'e coordinates, $P_{rr}$ leads to Equation~\ref{eqn:fallerdistance} as before.

In fact the radar metric is simply a special case of the spatial projector, or vice versa the spatial projector is the fully covariant generalisation of the radar metric. To see this, compute $P_{\mu\nu}$ for the comoving observer $u^\mu = \pm((-g_{00})^{-1/2},0,0,0)$. This matches the radar metric if Equation~\ref{eqn:radarmetric} is reinterpreted as $4$-dimensional, which amounts to padding the matrix of components with zeroes. Since the components agree within one coordinate system (any comoving one), they must agree in all coordinate systems, if the radar metric is to transform as a tensor.

A major advantage of the new line element is the faller's radial distance is clear from inspection (Equation~\ref{eqn:fallerdistance}). This provides much-needed contrast with the static measurement which is clear from inspection of Schwarzschild coordinates. Some of the most esteemed coordinates are null, so not insightful in this way. For other coordinate systems with $x^0$ timelike, including Kerr-Schild and Novikov coordinates, either the corresponding observers or their measurement are not clear from inspection. One might protest computations can be performed in any coordinates, which of course is true, however history shows this has not been sufficiently achieved in practice in this context. In fact, Schwarzschild coordinates could easily be misapplied to give
\begin{equation}
\label{eqn:mixeddistance}
ds = \big\vert e\big\vert\Schw^{-1}dr
\end{equation} 
for the faller's measurement. To arrive at this, express the falling observer's spatial metric in Schwarzschild coordinates, either by evaluating the projector directly, or transforming the radar metric from Lema\^itre coordinates. Now the correct radial vector to contract over is Equation~\ref{eqn:rvectorGP}, as this lies in the observer's local $3$-space; it leads to $dr/|e|$ as before. The mistaken choice is the Schwarzschild coordinate basis vector $\partial_r$, which is not orthogonal to $\mathbf u$. Contracting over this vector anyway picks out the component $P_{rr} = e^2(1-2M/r)^{-2}$, hence Equation~\ref{eqn:mixeddistance}. In fact this quantity is salvageable under a different physical interpretation: it relates a falling ruler to the coordinate gradient $dr$, but as determined within the \emph{static frame}.\footnote{By ``ruler'' we mean technically a vector orthogonal to $\mathbf u$ in the local tangent space, but intended as an approximation to an extended object on the manifold. This could be a hypothetical construction based on radar results, or a ``resilient'' physical rod \citep[\S2.5]{rindler1977} whenever the \emph{rod hypothesis} is justified.} By contrast Equation~\ref{eqn:fallerdistance} relates a falling ruler to $dr$ within its own \emph{falling frame}. Equation~\ref{eqn:staticdistance} relates a static ruler to $dr$ as determined within \emph{any frame}, because static particles remain at fixed $r$-values so foliation cannot alter this association. We will illustrate this in future work. Note the three measurements are consistent with the usual length-contraction formula, where the local Lorentz factor between the static and falling frames is:
\begin{equation}
\label{eqn:Lorentzfactor}
\gamma = -\mathbf u\cdot\mathbf u_\textrm{static} = |e|(1-2M/r)^{-1/2}
\end{equation}
However there are pitfalls and conceptual challenges when passing from the first-principles $\int ds$ proper distance or $1/\gamma$ length-contraction formulae to physical measurements.

The results are $1$-volume elements, giving the ratio of proper distance to coordinate gradient. Much pedagogy sets up a false dichotomy that $dr$ is not \emph{the} distance but $(1-2M/r)^{-1/2}dr$ \emph{is}. For example Equation~\ref{eqn:staticdistance} is termed ``radial ruler distance'' \citep[\S11.2]{rindler2006} or ``actual radial distance between two radial coordinates $r_A$ and $r_B$'' \citep[\S9]{moore2012}. These lack qualification as to which local frames do the measuring, or the intuition behind the foliation choice $t=\textrm{const}$. Conversely, others claim ``$r$ is not the radial distance'' \citep[\S9.7]{hobson+2006}, and similarly that the isotropic coordinate $r_\textrm{iso}$ is not radial distance. However we can find frames within which $r$ or $r_\textrm{iso}$ are exactly the radial distance. Better descriptions include the $dt=0$ slice as ``the correct spatial distance in the three-dimensional space defined by the static Killing vector'' \citep{senovilla2007}, and that Schwarzschild coordinates ``are adapted to observers at rest'' \citep[\S7.14]{hamilton2015}. Also, while there are entire books on relativistic measurement --- in theoretical physics \citep[\S9]{defeliceclarke1990} \citep{defelicebini2010}, astrometry \citep{soffel1989} \citep{kopeikin+2011}, and philosophy \citep{brown2005} --- our results fill an independent niche.

In forthcoming work, we use orthonormal frames to extend the formulae to angular momentum. What these various approaches to distance have in common is they measure within the local $3$-space orthogonal to the observer. Radar is not without its own limitations \citep{bini+2005}, in fact locally the approaches here give identical results. Some suggest the coordinate dependency on $r$ be removed, but should then the usual quantity $(1-2M/r)^{-1/2}dr$ be excised? There are certainly quantities defined independently of coordinates, such as the expansion tensor, or strain as predicted by gravitational waves. These are intrinsic properties, but often an external standard of reference is useful, and coordinates are literally such a ``map'' (chart/atlas) of spacetime. Finally, the lesson from introductory special relativity remains valid, that distance is relative to the observer's motion.

\section{Space and time coordinates}
\label{sec:spacetimecoords}

As is well known Schwarzschild $t$ becomes spacelike inside the horizon, and $r$ timelike. Hence one might question the use of a timelike coordinate to describe spatial distance in Equation~\ref{eqn:fallerdistance}. This section clarifies various properties of coordinate vectors and coordinate hypersurfaces for non-diagonal line elements, and ties up some remaining questions about distance measurement.

It is common to state space and time swap roles at the horizon, but this is also rightly criticised as a Schwarzschild coordinate property: ``Space and time themselves do not interchange roles: Coordinates do.'' \citep[\S3.7]{taylorwheeler2000} Recall the nature of a coordinate $\Phi$ is based on its level sets $\Phi=\textrm{const}$, a definition implied in various sources but rarely stated explicitly. We say $\Phi$ is timelike/null/spacelike when its orthogonal hypersurfaces are spacelike/null/timelike, meaning they have timelike/null/spacelike normals respectively. The gradient $d\Phi$ is one such normal $1$-form, with dual $(d\Phi)^\sharp$ a normal vector. Its nature is given by the sign of the squared-norm $(d\Phi)^\sharp\cdot(d\Phi)^\sharp = d\Phi\cdot d\Phi = g^{\Phi\Phi}$, which is simply a component of the inverse metric. Table~\ref{tab:coordnature} summarises this for the present coordinates.

\begin{table}[h]
\label{tab:coordnature}
\centering
\begin{tabular}{|c|c|p{5cm}|}
\hline
Coordinate hypersurface & Squared-norm of normal & Interpretation for\newline $r>2M$ / $r=2M$ / $r<2M$ \\
\hline
Schwarzschild $t$ & $g^{tt} = -\schw^{-1}$ & timelike / undefined / spacelike \\
Gullstrand-Painlev\'e $T$ & $g^{TT} = -1$ & timelike everywhere \\
$r$ & $g^{rr} = 1-\frac{2M}{r}$ & spacelike / null / timelike \\
Lema\^itre $\rho$ & $g^{\rho\rho} = \frac{e^2}{e^2-1+\frac{2M}{r}}$ & spacelike everywhere \\
$\theta$ & $g^{\theta\theta} = \frac{1}{r^2}$ & spacelike everywhere \\
$\phi$ & $g^{\phi\phi} = \frac{1}{r^2\sin^2\theta}$ & spacelike everywhere \\
\hline
\end{tabular}
\caption{The nature of coordinate hypersurfaces. The Gullstrand-Painlev\'e coordinate system has two timelike coordinates inside the horizon. The Lema\^itre coordinate system has consistent nature everywhere.}
\end{table}

However the nature of \emph{coordinate vectors} is distinct, in general. These are elements of the coordinate basis and have components $(\partial_\Phi)^{\mu} = \delta^\mu_\Phi$ for given $\Phi$. Hence each vector has squared norm $\partial_\Phi\cdot\partial_\Phi = g_{\Phi\Phi}$, so its nature is given by the sign of this metric component. If the metric is diagonal in a given coordinate system, then $g_{\Phi\Phi}=(g^{\Phi\Phi})^{-1}$ and so each coordinate vector has the same nature as the coordinate hypersurfaces. But in non-diagonal line elements there is room for misconception. See Table~\ref{tab:coordvectornature}. If sticking to purely geometric properties, the most one can state is all Killing vector fields are spacelike inside the horizon, and in particular the Killing vector field which is timelike at infinity becomes spacelike inside the horizon.

\begin{table}[h]
\label{tab:coordvectornature}
\centering
\begin{tabular}{|c|c|p{5cm}|}
\hline
Coordinate vector & Squared norm & Interpretation for\newline $r>2M$ / $r=2M$ / $r<2M$ \\
\hline
Schwarzschild $\partial_t$ & $g_{tt} = -\schw$ & timelike / null / spacelike \\
Gullstrand-Painlev\'e $\partial_T$ & $g_{TT}$ = $-\frac{1}{e^2}\schw$ & as above \\
Lema\^itre $\partial_T$ & $g_{TT} = -1$ & timelike everywhere \\
Schwarzschild $\partial_r$ & $g_{rr} = \schw^{-1}$ & spacelike / null / timelike \\
Gullstrand-Painlev\'e $\partial_r$ & $g_{rr}$ = $\frac{1}{e^2}$ & spacelike everywhere \\
Lema\^itre $\partial_\rho$ & $g_{\rho\rho} = \frac{1}{e^2}\Big(e^2-1+\frac{2M}{r}\Big)$ & spacelike everywhere \\
\hline
\end{tabular}
\caption{The nature of coordinate basis vectors. For Gullstrand-Painlev\'e coordinates all basis vectors are spacelike inside the horizon. Omitted are $\partial_\theta$ and $\partial_\phi$, which have the same nature as the gradients $d\theta$ and $d\phi$ above, because all our line elements are diagonal in $\theta$ and $\phi$.}
\end{table}

A surprising feature of Table~\ref{tab:coordvectornature} is that $\partial_r$ for Gullstrand-Painlev\'e coordinates is a distinct vector from $\partial_r$ in Schwarzschild coordinates. This is despite the $r$-coordinate being identical in both cases --- in the sense of a scalar field which matches on chart overlaps, and both having the same components $(0,1,0,0)$ in their respective systems. The standard vector transformation law yields:
\begin{equation}
\label{eqn:rvectorGP}
\partial_r^\textrm{(GP)} = \pm\frac{1}{e}\Schw^{-1}\eroot\partial_t^\textrm{(Schw)} + \partial_r^\textrm{(Schw)}
\end{equation}
There are few mentions of this potential error-causing subtlety in the literature \citep[\S1]{finch2015} \citep{bini+2012}, although there are related comments about each basis vector depending on \emph{every} element of the dual basis \citep[\S3.3]{schutz2009}, or ambiguities with partial derivative notation. Recall a basis and its dual are related by
\begin{equation}
\label{eqn:dualbasis}
dx^\mu(\partial_\nu) = \delta^\mu_\nu
\end{equation}
A coordinate dual basis element $d\Phi$ depends only on the coordinate $\Phi$, since it is related to the level sets $\Phi=\textrm{const}$. This is not the case for the vector $\partial_\Phi$, as without specifying the accompanying coordinates all one can say is $d\Phi(\partial_\Phi)=1$. Note duality of bases is distinct from duality of individual vectors: $(d\Phi)^\sharp\ne\partial_\Phi$ in general. We could instead define a unique coordinate vector depending only on $\Phi$:
\begin{equation}
\partial_\Phi^\textrm{(unique)} := \frac{(d\Phi)^\sharp}{d\Phi\cdot d\Phi}
\end{equation}
which is orthogonal to the level sets $\Phi=\textrm{const}$ and satisfies $d\Phi(\partial_\Phi^\textrm{(unique)}=1$. For diagonal line elements it coincides with the default coordinate vector. However in our context the flexibility of $\partial_r$ is actually helpful! Both spatial measurement and Einstein simultaneity require vectors orthogonal to an observer. From Equation~\ref{eqn:dualbasis}, $\partial_r^\textrm{(GP)}$ is orthogonal to $dT$, so this vector lies in the $3$-space of a falling observer. Similarly $\partial_r^\textrm{(Schw)}$ is orthogonal to $dt$, so it lies in the $3$-space of a static observer. This justifies the naive contractions in Section~\ref{sec:spatialdistance}. Our well-suited coordinates glossed over this requirement automatically. This also partly answers the earlier question: even though $r$ is timelike inside the horizon, the measurement direction $\partial_r^{(GP)}$ is spacelike.

So why does the proper distance $dr/|e|$ diverge as $e\rightarrow 0$? This is because the $r$-coordinate, which is timelike for $r<2M$, is \emph{purely} timelike to the zero Killing energy observers, meaning $(dr)^\sharp$ is parallel to the $4$-velocity $\mathbf u$. In an observer's frame, distances are measured along vectors $\mathbf v$ orthogonal to $\mathbf u$, but for $e=0$ observers the gradient of $r$ is zero in all these directions: $dr(\mathbf v)=0$, so cannot describe any spatial measurement. This situation is much less exotic than it appears. The same thing occurs for static observers outside the horizon, for whom the $t$-coordinate is purely timelike so cannot describe space. In fact this situation occurs for spacelike coordinates also, if they are orthogonal to the ruler direction: none of our observers can describe \emph{radial} distance in terms of the $\theta$ or $\phi$-coordinates! Returning to the $e=0$ observers, one can set $dr=0$ in Equation~\ref{eqn:Schwcoords} to obtain the radial proper distance $\sqrt{2M/r-1}\,dt$ in terms of the $t$-coordinate gradient.

Our proper distance expressions are only defined along the $1$-dimensional ruler direction. If $dr/|e|$ were instead interpreted as a $1$-form on an entire $4$-dimensional tangent space, indeed this would be a timelike $1$-form inside the horizon. This would mean the vector $(dr/|e|)^\sharp$, the direction of steepest gradient, is timelike. But when restricted to the $1$-dimensional subspace of the tangent space parallel to the ruler, $dr/|e|$ and its dual are spatial.


\section{Simultaneity, and time at infinity}
\label{sec:simultaneity}

Schwarzschild $t$ is called the ``time at infinity''. At infinity this is unambiguously true, as the proper time for a static observer is $d\tau = \sqrt{1-2M/r}\,dt \rightarrow dt$ as $r\rightarrow\infty$. However at events with $r<\infty$, the interpretation of their $t$-coordinate as the time at infinity \emph{right now} makes an implicit simultaneity assumption. This is justified as one of the most natural choices, based on the static Killing vector field. But the falling observers lead to a different convention based on the hypersurfaces orthogonal to the congruence. Under this alternative, the time passing at infinity is only \emph{finite} for a falling test particle to cross $r=2M$. This goes beyond the usual statement that the falling particle's proper time is finite, to a simultaneity convention extended across spacetime. This alternate convention is conceptually motivated, helps avoid common misconceptions, and may have practical results for calculations which depend on foliation.

Recall in general relativity, any spatial hypersurface can be considered a ``simultaneous'' instant of time. Some authors allow the hypersurface to be null or even go timelike in places. In symmetric and physically relevant spacetimes, one would expect to pick out natural choices from amongst the infinitude of options. For example in static spacetimes, the hypersurfaces orthogonal to the timelike Killing vector make a natural choice of simultaneity. In spacetimes with a vorticity-free congruence, the hypersurfaces orthogonal to it make a natural choice, as is the approach in cosmology. (Orthogonality to the observer is a consequence of Einstein-Poincar\'e simultaneity.)

\begin{figure}
\label{fig:simul}
\centering
\includegraphics[width=0.7\textwidth]{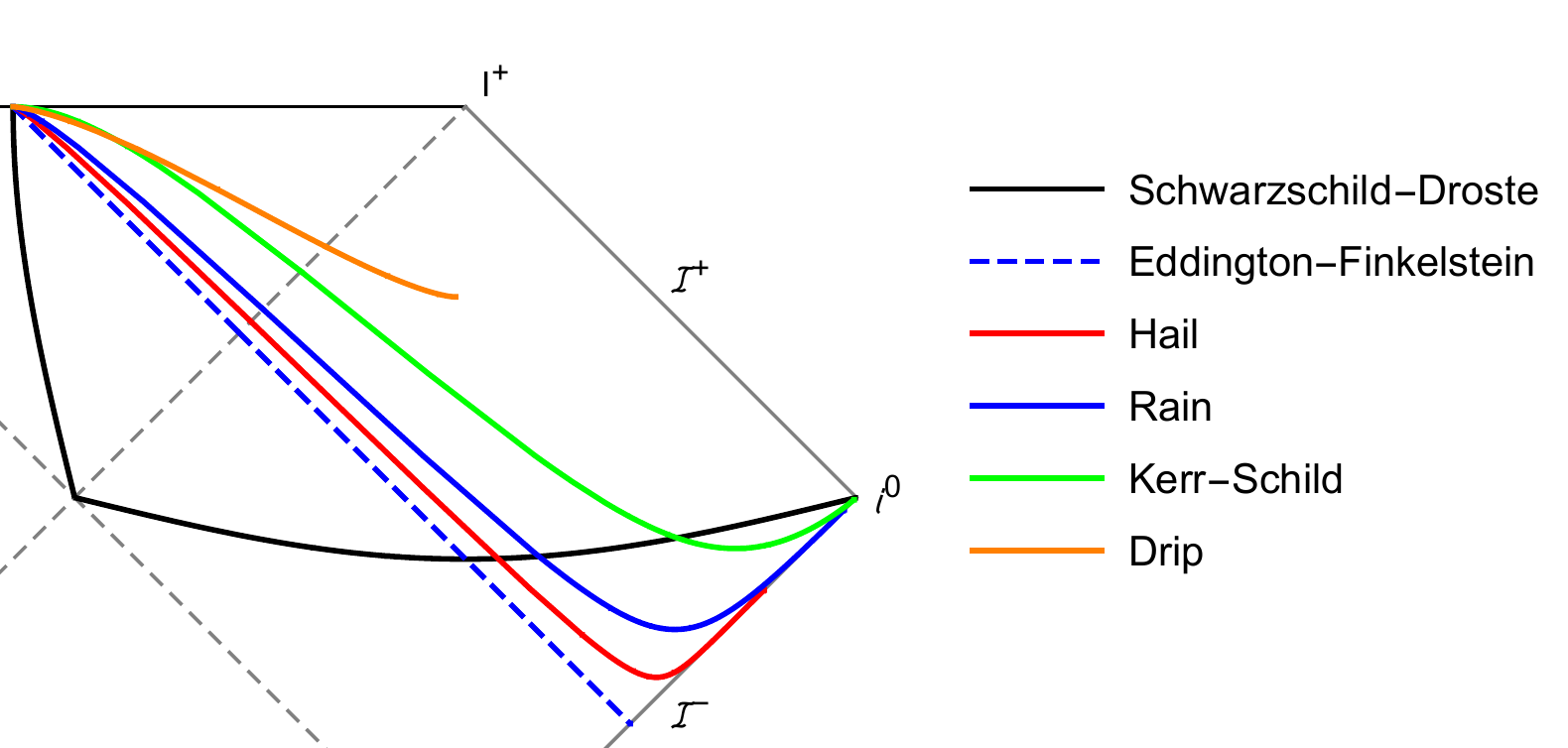}
\caption{Simultaneity conventions defined by $x^0=\textrm{const}$ in various coordinate systems, in the physical regions I and II. Each line shows the events considered simultaneous with a given event near $r=0$. The rain, hail, and drip hypersurfaces are orthogonal to the worldlines in Figure~\ref{fig:rainhaildripmist}, and for $r>2M$ the Schwarzschild choice is orthogonal to static observers. Those simultaneity choices reach to spatial infinity, but with different angles of approach. The drip hypersurface extends only to $r_\textrm{max}$, and the Eddington-Finkelstein coordinate is null.}
\end{figure}

Figure~\ref{fig:simul} shows some possible conventions for Schwarzschild. Note the drip, rain, and hail simultaneity curves appear from top to bottom in that order. This means the events considered simultaneous to the given $r\approx 0$ event are earlier for hail than for rain, for example. Conversely, for a fixed location $r_0$ say, simultaneous events at $r<r_0$ are later for hail than rain, so events at $r<r_0$ occur earlier under hail simultaneity; vice versa for $r>r_0$. This is consistent with simultaneity in special relativity, as applied to each local frame. Recall that for a Lorentz-boosted frame, events ahead of them (in the $3$-direction of the boost) occur earlier than they do for the unboosted frame, and vice versa. These statements of causal order are not absolute, even though we seek natural choices. The only absolute standard of causality is based on the light cone.

In Schwarzschild, consider a radially falling test particle. Its $t$-coordinate diverges as it passes the horizon, then in the black hole interior $t$ \emph{decreases} towards the future, at least for $e>0$ (Equation~\ref{eqn:velocity}). Historically this led to much confusion, but modern pedagogy stresses the proper time is finite and the divergence at $r=2M$ is merely a Schwarzschild coordinate issue. However, it is still said $t$ is the time at infinity, rarely with any qualifiers. But how would this view deal with the old misconception that time runs backwards in the black hole interior? One might then limit the interpretation of $t$ as time at infinity to region I. But the pathology at $r\rightarrow 2M$ remains: would a distant observer say the particle freezes at the horizon, or that black holes never form? It seems observers at infinity still hold to pre-1960s views. Note we do not mean the visual appearance of the observer, which fades exponentially as is well-known \citep[\S32.3]{misner+1973}, as simultaneity is not determined only by when null signals reach an observer, but also accounts for travel time of the signals \citep{scherr+2002}. Note also $t$ does not coincide with the arrival time of photons at distant $r$, which is described by the outgoing Eddington-Finkelstein null coordinate which has only qualitatively similar behaviour to $t$ for the falling particle.

There is a \emph{local} physical interpretation for $t$. For a static observer, it is their proper time divided by the redshift factor $\sqrt{\boldsymbol\xi\cdot\boldsymbol\xi}$ to compensate for gravitational time-dilation. Relying on the time-symmetry of their worldlines through spacetime, static observers can determine their relative time-dilation factors by signalling one another. We can even find a physical interpretation of $t$ for the falling particle. Consider a line of static observers along the faller's worldline, where each records the time interval for the faller to pass them. The particle crosses an interval $dr$ in proper time $d\tau = dr/u^r$, but according to the local static frame the faller is time-dilated so the actual time passed is increased by the Lorentz factor between the frames (Equation~\ref{eqn:Lorentzfactor}). Now if the static observer were to further increase this quantity to compensate for their gravitational time-dilation as above, the resulting ``time at infinity'' would be $dr\cdot u^t/u^r$ which is $dt$. Note other frames give different results, and this interpretation relies on static frames which are impossible for $r\le 2M$ and physically unreasonable at $2M+\epsilon$ due to extreme acceleration. Compare the discussion of Equation~\ref{eqn:mixeddistance}.

Simultaneity defined from the faller congruence is more physically realistic, and extends inside the horizon. Locally, a falling test particle has coordinate time matching its proper time: $dT = d\tau$. Under the new simultaneity convention, the same coordinate interval $dT$ passes everywhere. At infinity (which requires $e\ge 1$), this corresponds to a proper time $dT/e$ for a static observer. Hence we conclude the time at infinity is $T/e$. In particular, the time at infinity for the particle to cross the horizon is finite! This avoids misconceptions such as the particle freezing at the horizon, black holes never forming, one having unlimited time to fly down and rescue a falling astronaut, and so on. The $e=1$ convention is especially natural.


\begin{figure}
\label{fig:zerosimul}
\centering
\includegraphics[width=0.5\textwidth]{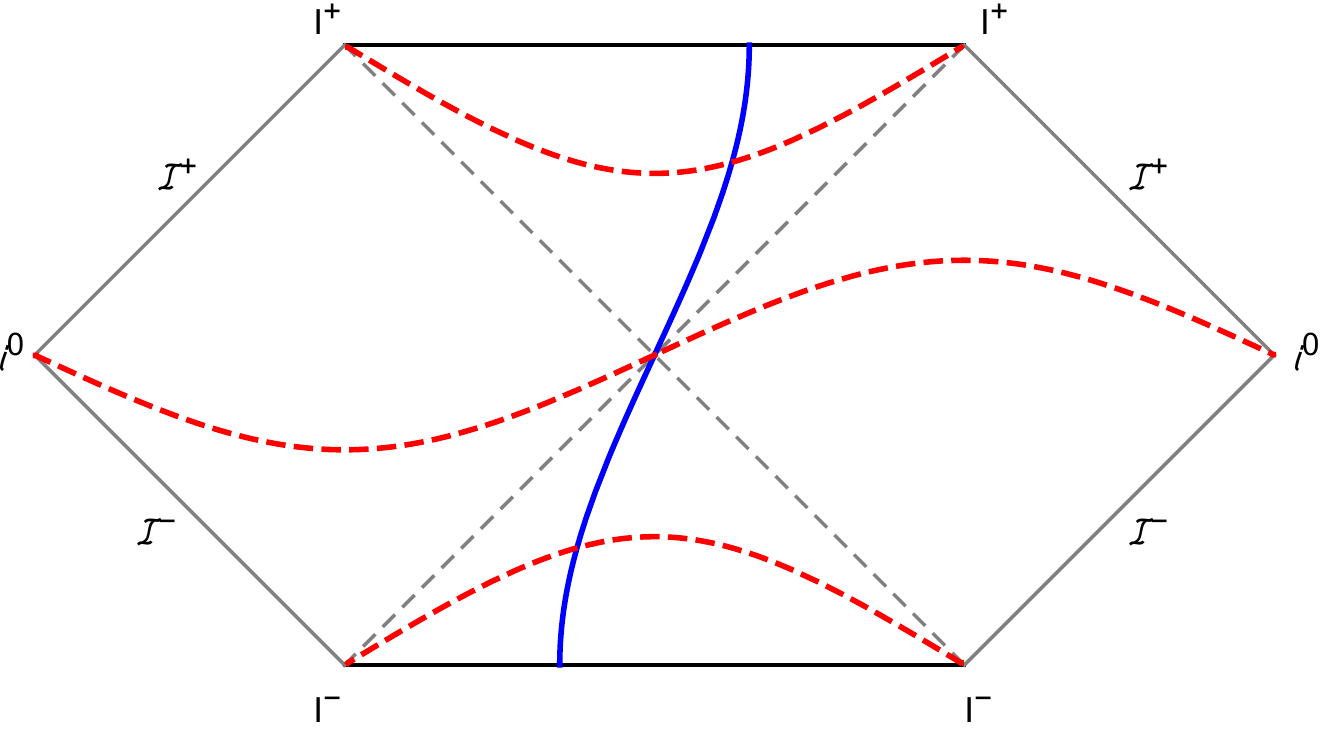}
\caption{Simultaneity for the zero Killing energy congruence. A representative worldline $t=\textrm{const}$ is shown (blue), and orthogonal hypersurfaces (red) at three selected events. The upper and lower curves are $r=\textrm{const}$. At the bifurcate horizon, the orthogonal hypersurface leaves the congruence but can be extended geodesically into the exterior regions, where it concurs with static simultaneity.}
\end{figure}

The $e=0$ congruence leads to qualitatively different simultaneity. The orthogonal hypersurfaces $r=\textrm{const}$ are the infinite $3$-cylinders mentioned in Section~\ref{sec:embedding}, except for $r=2M$ as Figure~\ref{fig:zerosimul} shows. Incidentally, one might wonder what the ``radial'' direction is for these observers. The remaining spatial direction is orthogonal to $\partial_\theta$ and $\partial_\phi$, hence must be $\partial_t$. However in the $3$-dimensional space, $\partial_t$ points along the axis of the cylinder $\mathbb R\times S^2$, so it is better conceived as a translation vector not radial (it turns out others have also mentioned this). Likewise in $4$ dimensions one thinks of $\partial_t$ as translational, at least in regions I and III where it is translation in time.

Typical pedagogy can leave the impression of a single global time slicing, in certain aspects. An exception is \citet[\S3.2.2]{frolovnovikov1998}, who make clear the existence of different choices. They advocate the usual $t$-slices, rejecting the ``Lema\^itre frame'' ($e=1$) because it is not a rigid congruence. However one would not expect rigidity for a freefalling congruence, as tidal forces are a sign of gravity.

\section{Conclusion}
The coordinates and foliation induced by timelike radial geodesic observers provide an insightful contrast to Schwarzschild coordinates and the static foliation. We examined this contrast for the curvature of $3$-dimensional space and its embedding diagram, measurement of radial proper distance, simultaneity and the time at infinity, as well as other areas. A special case of these radial geodesics and their accompanying foliation has already proved useful for Hawking radiation and analogue gravity, and we anticipate further applications in these areas in the general case.





\bibliographystyle{hapj}
\bibliography{biblio}


\subsection*{Acknowledgment}
Tamara Davis set me on the course of conceptual issues in GR. Discussions with Geraint Lewis, Charles Hellaby, Tehani Finch, Eric Poisson, and others were also insightful.

\end{document}